\documentclass[fleqn,10pt]{wlscirep}
\usepackage[utf8]{inputenc}
\usepackage[T1]{fontenc}
\usepackage{siunitx}
\usepackage{threeparttable,booktabs}

\usepackage{bm}
\newcommand{\euler}{\mathrm{e}}
\newcommand{\imag}{\mathrm{i}}
\newcommand{\vect}[1]{\bm{{#1}}}
\newcommand{\vects}[1]{\bm{#1}}
\newcommand{\lapl}{\nabla^2}
\newcommand{\para}[1]{\left( #1 \right)}
\newcommand{\grad}{\vects{\nabla}}

\newcommand{\bvis}{\eta_{\mathrm{B}}}
\newcommand{\bkt}[1]{\left[ #1 \right]}
\newcommand{\avg}[1]{\left \langle #1 \right \rangle}

\newcommand{\velpot}{\varphi}

\newcommand{\na}{\mathrm{a}}

\newcommand{\cfl}{c_0}

\newcommand{\bcdot}{\bm{\cdot}}
\newcommand{\divg}{\vects{\nabla}\bcdot}

\title{Measuring and simulating the biophysical basis of the acoustic contrast factor of biological cells}

\author[1,2,3,+,*]{Cooper Lars Harshbarger}
\author[3,+]{Alen Pavlic}
\author[1,2]{Davide Cesare Bernardoni}
\author[3]{Amelie Viol}
\author[1,2]{Jess Gerrit Snedeker}
\author[3${\dagger}$]{Jürg Dual}
\author[1,2,4,5${\dagger}$]{Unai Silv\'an}
\affil[1]{Department of Orthopedics, Balgrist University Hospital, University of Zurich, Zurich, Switzerland}
\affil[2]{Institute for Biomechanics, Swiss Federal Institute of Technology Zurich, Zurich, Switzerland}
\affil[3]{Institute for Mechanical Systems, Swiss Federal Institute of Technology Zurich, Zurich, Switzerland}
\affil[4]{BCMaterials, Basque Center for Materials, Applications and Nanostructures, Leioa, Spain}
\affil[5]{Ikerbasque, Basque Foundation for Science, Bilbao, Spain}

\affil[*]{hcooper@ethz.ch}

\affil[+]{these authors contributed equally to this work}
\affil[$\dagger$]{these authors contributed equally to this work}


\begin{abstract}
The acoustic contrast factor (ACF) is calculated from the relative density and compressibility differences between a fluid and an object in the fluid. To name but one application, this acoustic contrast can be exploited using acoustophoretic systems to isolate cancer cells from a liquid biopsy, such as a blood sample. Knowing the ACF of a cancer cell represents a crucial step in the design of acoustophoretic systems for this purpose, potentially allowing the isolation of circulating cancer cells without labels or contact. For biological cells the static compressibility is different from the high frequency counterpart relevant for the ACF. In this study, we started by characterizing the ACF of low vs. high metastatic cell lines with known associated differences in phenotypic static E-modulus. The change in the static E-modulus, however, was not reflected in a change of the ACF, prompting a more in depth analysis of the influences on the ACF. We demonstrate that static E-modulus increased biological cells through formaldehyde fixation have an increased ACF. Furthermore, the static E-modulus decreased biological cells treated with actin polymerization inhibitor cytochalasin D have a decreased ACF. Complementing these mechanical tests, a numerical COMSOL model was implemented and used to parametrically explore the effects of cell density, cell density ratios, dynamic compressibility and therefore the dynamic bulk modulus. Collectively the combined laboratory and numerical experiments reveal that a change in the static E-modulus alone might, but does not automatically lead to a change of the dynamic ACF for biological cells. This highlights the need for a multiparametic view of the biophysical basis of the cellular ACF, as well as the challenges in harnessing acoustophoretic systems to isolate circulating cells based on their mechanical properties alone.
\end{abstract}
\begin{document}

\flushbottom
\maketitle

\thispagestyle{empty}

\section*{Introduction}

\begin{figure}[h!]
    \centering
    \includegraphics[width=\linewidth]{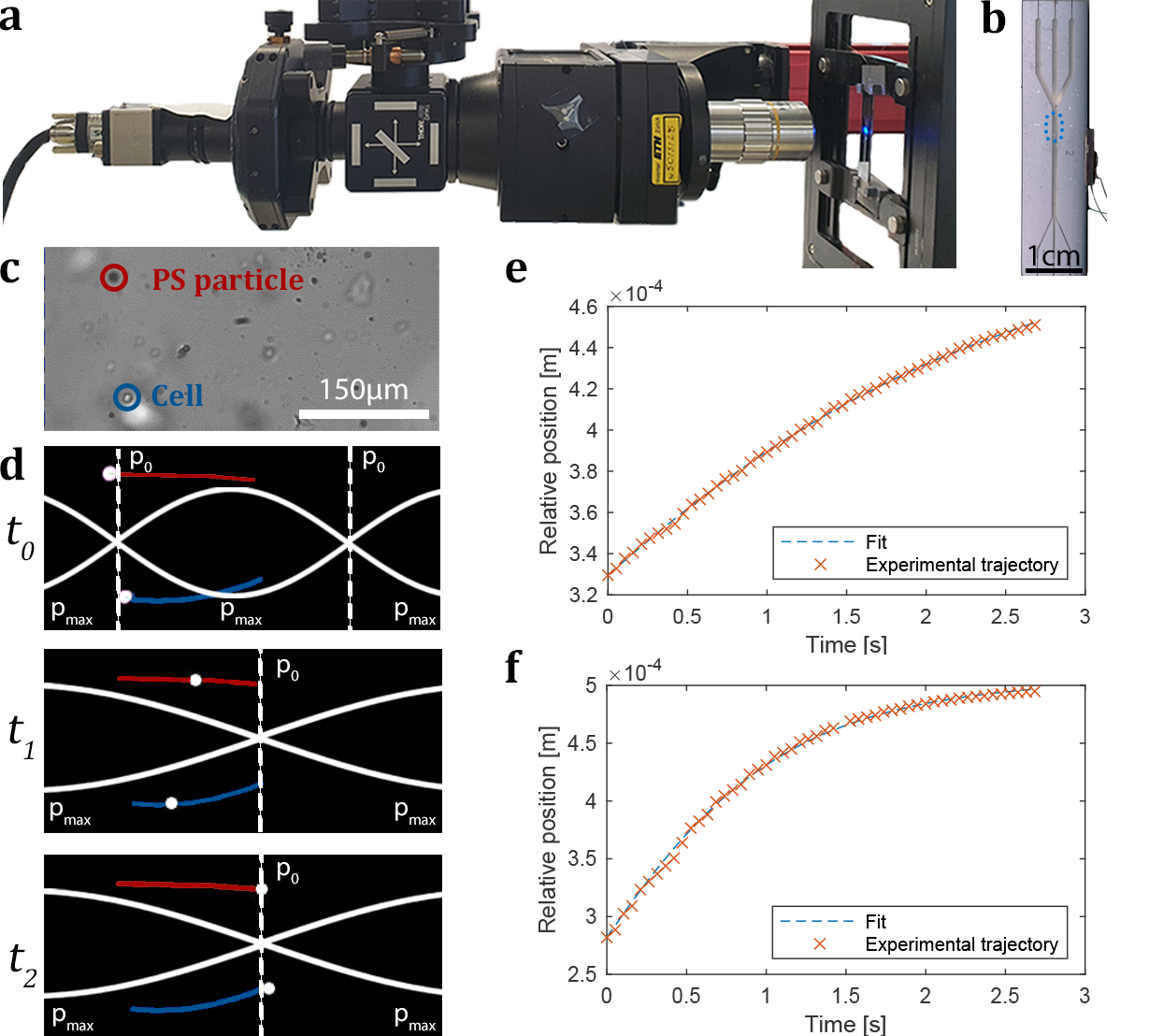}
    \caption[Setup and measurement protocol of the ACF of a biological cell.]{(\textbf{a}) Image of the vertical setup showing the camera (far left) and the AF device being imaged (far right). (\textbf{b}) AF device imaged with the ROI indicated by the blue dashed box. A piezoelectric transducer glued to the side or bottom of the device establishes an acoustic field within the device. (\textbf{c}) Frame of a video of a cell and PS particle in suspension within the AF device. Wherever possible, the cells were pre-aligned in a $\lambda$ mode. (\textbf{d}) The analysis of the particle motion is performed in multiple steps: The video recording of every experiment is split into single images and converted to black and white. The object trajectories are indicated in the same colors as the PS particle and cell in \textbf{c} and are generated using the Fiji Plugin TrackMate. At $t_0$ and $\SI{1.42}{\mega\hertz}\pm\SI{2}{\kilo\hertz}$ the cells and PS particle are located in pressure nodes of the $\lambda$ mode where $p_0 = 0$, as both objects have a positive acoustic contrast factor. The frequency is changed to $\SI{742.1}{\kilo\hertz}\pm\SI{2}{\kilo\hertz}$, which alters the location of the zero pressure nodes. This causes the PS particles and biological cells to migrate during $t_1$ away from their previous position to the middle of the channel. At $t_2$ the objects are again at the zero pressure nodes $p_0 = 0$ of the $\lambda/2$ mode, thus having minimized their acoustic potential. (\textbf{e}, \textbf{f}) The plots show the position of a biological cell (\textbf{e}) and PS (\textbf{f}) particle during the migration from a high pressure to a low pressure region of the $\lambda/2$ mode. The fit is with respect to the pressure within the system using eq. \ref{equation_trajectory}. Using a self written Matlab script, the ACF of the biological cells can be back-calculated.}
    \label{fig:setup}
\end{figure}

Carl Jung states "The greater the contrast, the greater the potential."\cite{alma990005599940205503}. This is true in the field of acoustofluidics (AF), the frequency dependent acoustic pressure wave driven manipulation of particles in a fluid. In the case of AF, the contrast is termed the acoustic contrast factor (ACF) $\Phi$, and arises from relative density and compressibility differences between a fluid and an object in the fluid as described by Yosioka and Kawasima\cite{yosioka1955acoustic}. The greater the ACF, the greater the acoustic potential and the more the object in the fluid will interact with the pressure wave. Changes to either the object or the fluid will lead to a change of potential, whereas the acoustic potential in an AF system is termed the Gor'kov potential\cite{gor1962forces}. Exploiting the acoustic contrast between biological cells and a fluid has been successfully demonstrated multiple times for the manipulation of cancer cells in suspension\cite{undvall2021two, augustsson2012microfluidic, cushing2018reducing, olm2019label, antfolk2017label}. Therefore, AF is a promising diagnostic tool when considering the spreading of cancer cells through bodily fluids, a process known as metastasis which is responsible for 90\% of cancer related deaths\cite{gupta2006cancer}. In order for cancer cells to enter the metastatic state, they undergo a wide range of changes, including a decrease in their static E-modulus\cite{cross2020nanomechanical}, which leads to a different mechanical phenotype between the cancer cells located within the primary tumor and the disseminated cancer cells. This decrease in the static E-modulus can be measured by gold standard measurement techniques such as magnetic tweezers\cite{alenghat2000analysis}, AFM\cite{holenstein2019relationship} and micropipette aspiration \cite{hochmuth2000micropipette}. Although the presented methods are powerful and established, there are certain limitations such as that the biological cells are adherent on tissue culture plastic, that only one mechanical parameter of the cell of interest is measured and that the measurement is based off of one small region of the biological cell. An alternative method to measure the static E-modulus, which relies on the deformation of the entire biological cell and does not rely on adherent cells, is real-time deformability cytometry (RT-DC)\cite{otto2015real}. Although these methods have led to a vast catalog of static E-moduli for various cancer cell types in adherent and non-adherent states, the reported material properties stem from (quasi-)static measurements. Considering emerging ultrasound frequency technologies to detect and treat cancer, such as oncotripsy\cite{heyden2016oncotripsy, schibber2020dynamical}, the question needs to be answered if static material parameters are sufficient or if lookup tables for material properties need to be extended with values stemming from dynamic measurements. Such dynamic material properties can be measured with AF, by focusing cells using ultrasound frequencies, as illustrated in Fig. \ref{fig:setup}. By focusing the biological cells in a suspension with polystyrene (PS) particles with a known ACF, the ACF of the biological cells can be calculated. This method of measuring the ACF is contactless, label-free and does not decrease cell viability\cite{wiklund2012acoustofluidics}.

\begin{figure}[h]
    \centering
    \includegraphics[width=\linewidth]{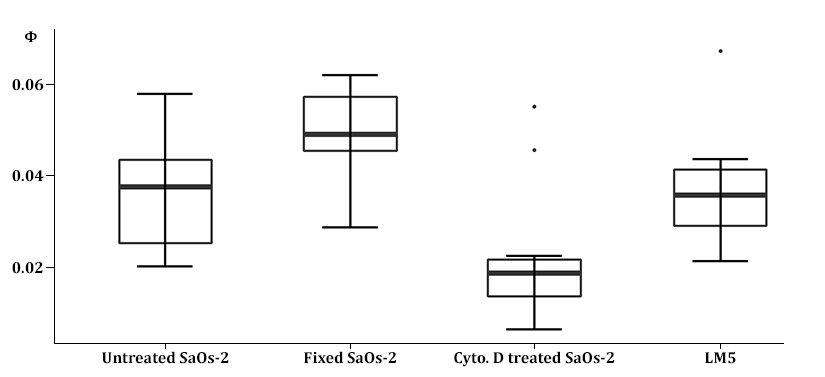}
    \caption[ACF comparison of cell lines and treatments.]{The ACF measured in PBS of the low metastatic potential cell line SaOs-2 is taken as the baseline and is 0.037 ($n = 25$). SaOs-2 formaldehyde fixed cells have an ACF of 0.050 ($n = 10$). SaOs-2 cells treated with cytochalasin D have an ACF of 0.022 ($n = 12$). The higher metastatic potential cell line LM5 have an ACF of 0.037 ($n = 12$).}
    \label{fig:phi_readout}
\end{figure}

There is a growing body of literature where the ACF of biological cells is measured\cite{hartono2011chip, wang2018single, cushing2017ultrasound, augustsson2010measuring} as summarized in table \ref{tab:eukaryote_mat_prop} and new ACF values reported in Fig. \ref{fig:phi_readout}. The work prior to this study however mostly results in measurements of the compressibility of biological cells from which the ACF can be calculated, whereas the density is assumed\cite{hartono2011chip}, measured with methods such as a neutrally buoyant sample \cite{cushing2017ultrasound} or measured by defocusing of the particles during sedimentation\cite{augustsson2010measuring}. Further publications resulted in lookup tables for cell lines with different metastatic potentials\cite{wang2018single} and ACF comparisons between fixed and non-treated biological cells\cite{cushing2017ultrasound}.

Beside these few studies, little is known about the correlation between the known static E-modulus of biological cells and the dynamic ACF. To help broaden the understanding of the applicability of known material parameters to ultrasound frequency applications we increased the static E-modulus\cite{pillarisetti2011mechanical} of the bone cancer cell Sarcoma Osteogenic (SaOs)-2 using the protein crosslinking fixative formaldehyde\cite{thavarajah2012chemical} and decreased static E-modulus\cite{wakatsuki2001effects} using the actin polymerization inhibitor cytochalasin D\cite{schliwa1982action} (CD). After these treatments the ACF of the biological cells are measured in a Phosphate Buffer Solution (PBS). Furthermore, the SaOs-2 parental cell line is paired with its highly metastatic counterpart Lung Metastasis (LM)5, with the known static E-modulus measured of $\SI{0.95}{\kilo\pascal}$ and $\SI{0.8}{\kilo\pascal}$, respectively\cite{holenstein2019relationship} when measured with RT-DC. Thus demonstrating the limitations of trying to predict the ACF when only taking a change in static E-modulus into account. To further place the experimental results into context, a numerical model, seen in Fig. \ref{fig:numerical_model} is introduced. The parameter space investigated with the model includes the density, compressibility and static E-modulus, as seen in Fig. \ref{fig:numerical_model_results}. Thus aiming at closing gaps of unknown dynamic material properties of cells which would be useful for cancer diagnosis and treatment. 


\section*{Results}

We used a vertical setup composed of a widefield microscope to image the PS particles and cell migrate through the fluid, as seen in Fig. \ref{fig:setup}. The AF device was placed in a vertical orientation to potentially eliminate rolling along the glass surface and the 3 inlet configuration can improve the prefocusing efficiency, if needed. The standing pressure wave was established by a piezoelectric transducer powered by a sinusoidal signal of a function generator increased in magnitude by an amplifier.  The frequencies used are in the $\SI{742.1}{\kilo\hertz}\pm\SI{2}{\kilo\hertz}$ range and were varied for optimal focusing. A summary of the ACFs of previous publications is, together with the results of this study, summarized in table \ref{tab:eukaryote_mat_prop}. The ACF measured in PBS of the low metastatic potential cell line SaOs-2 is taken as the baseline and is 0.037 ($n = 25$). SaOs-2 formaldehyde fixed cells have an ACF of 0.050 ($n = 10$). SaOs-2 cells treated with cytochalasin D have an ACF of 0.022 ($n = 12$). The higher metastatic potential cell line LM5 have an ACF of 0.037 ($n = 12$). The average cell diameter for each condition was measured and is reported in table \ref{tab:eukaryote_mat_prop}. 

\begin{figure}
    \centering
    \includegraphics[width=\linewidth]{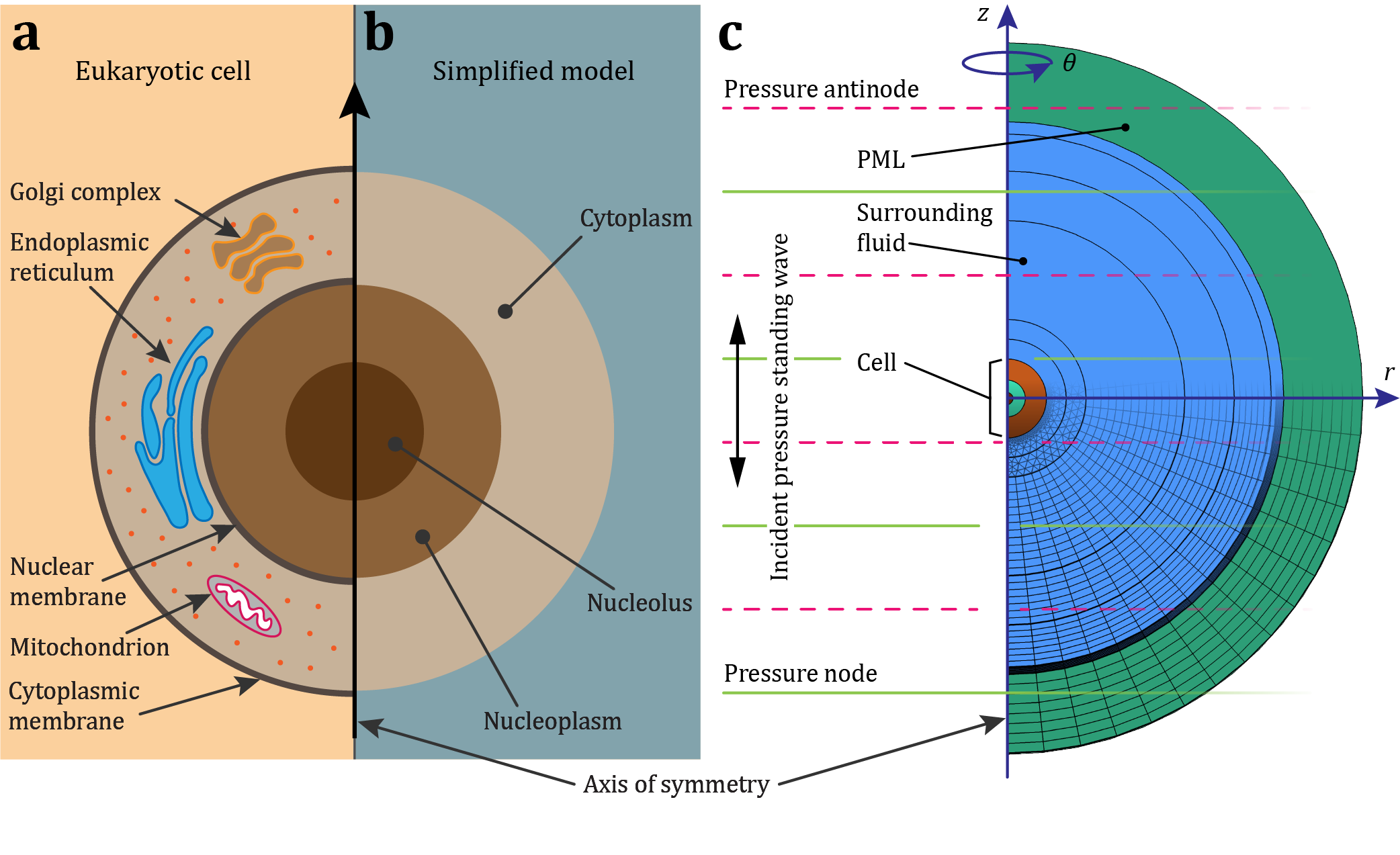}
    \caption[Numerical model of biological cell.]{Numerical model of a eukaryotic cell. The ACF and its dependencies are extracted from the numerical model. (\textbf{a}) The structure of a typical eukaryotic cell. (\textbf{b}) Simplification of the representation of the cell into three layers, namely, cytoplasm, nucleoplasm, and nucleolus. (\textbf{c}) Geometry of an axisymmetric finite-element method model of a eukaryotic cell, surrounded by fluid and perfectly-matched layer (PML).}
    \label{fig:numerical_model}
\end{figure}

The base numerical model of a SaOs-2 cell is constructed as described in the Methods, and is composed of a cytoplasm-nucleoplasm-nucleolus structure as shown in Fig. \ref{fig:numerical_model}. The cell and all the layers are assumed to be spherical. The radius for the SaOs-2 cell is taken from the measurements reported in table \ref{tab:eukaryote_mat_prop} ($\SI{8.80}{\micro\meter}$). The outer radius of the nucleoplasm ($\SI{6.66}{\micro\meter}$) is computed from previously reported\cite{holenstein2019relationship} nucleus-cell volume ratio of $0.434$, while the radius of the nucleolus ($\SI{2.14}{\micro\meter}$) follows from the nucleolus-nucleus volume ratio of $1/30$ reported by Guttman and Halpern\cite{guttman1935nuclear}. The mass density in each layer is computed according to the mass density ratios of the three characteristic cellular layers, as reported by Kim and Guck\cite{kim2020relative}, while imposing an assumed apparent density of the whole cell of $\SI{1060}{\kilo\gram\per\cubic\meter}$. The static E-modulus of each cellular layer is assigned based on the ratios used in the literature\cite{heyden2016oncotripsy}, such that the apparent static E-modulus computed through the rule of mixture results in $E=\SI{1}{\kilo\pascal}$, as reported from RT-DC measurements\cite{holenstein2019relationship}. The initial bulk modulus of individual cellular layer is assigned based on the reported 2\% lower speed of sound in nucleus compared to the cytoplasm\cite{taggart2007ultrasonic}.

\begin{figure}
    \centering
    \includegraphics[width=\linewidth]{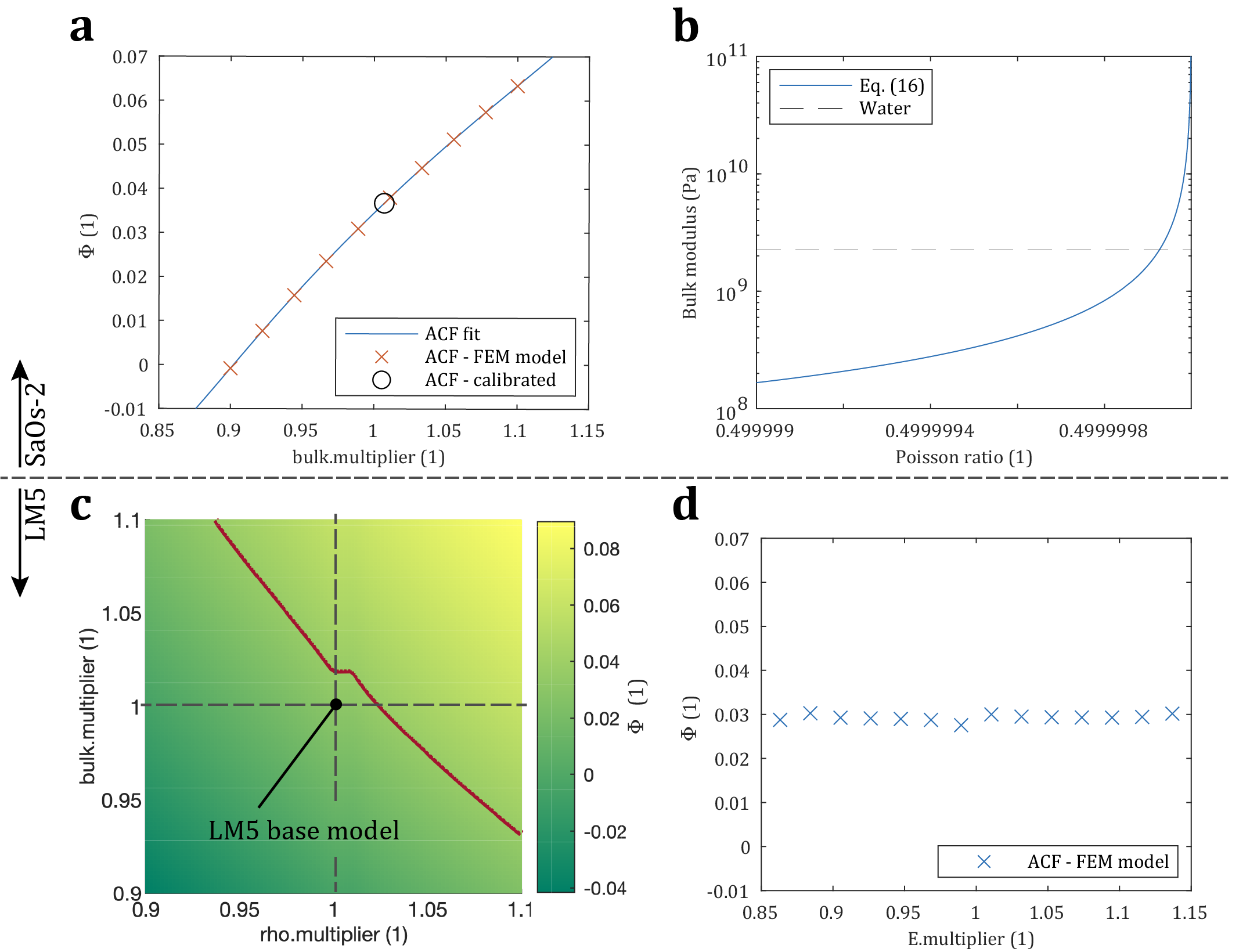}
    \caption[Results of the numerical model of a biological cell.]{(\textbf{a}) The ACF of a SaOs-2 cell depending on the bulk modulus multiplier (bulk.multiplier) common to the three layers of the SaOs-2 model with a constant static E-modulus. To calibrate the model, the bulk multiplier is tuned to the value that yields the ACF of $0.037$ measured experimentally. (\textbf{b}) The Poisson's ratio corresponding to the range of bulk modulus relevant for our study, computed from eq. \ref{eq:linear_elastic} with a constant static E-modulus; the bulk modulus of water is shown for reference. (\textbf{c}) The calibrated layer-based material properties from the SaOs-2 model were used as a basis for the LM5 model, changing only the volume of the cell and the nucleus-cell volume ratio according to experimental observations\cite{holenstein2019relationship}. The base LM5 model underestimates the ACF at $0.0309$, compared to the experimentally measured value of $0.037$. The common bulk modulus multiplier and density multiplier are varied in (\textbf{c}) to indicate the combination changes in the density and bulk modulus that provide the experimentally measured ACF (denoted with the red line). (\textbf{d}) Varying the E-modulus, while keeping the bulk modulus and the density constant, does not influence the ACF.}
    \label{fig:numerical_model_results}
\end{figure}

In the first step shown in Fig. \ref{fig:numerical_model_results}(a), the SaOs-2 model, with the density and static E-Modulus taken from literature, is calibrated by varying the bulk modulus $K$ of all the layers through a common multiplier to match the experimentally measured ACF of $0.037$. The bulk modulus of the calibrated SaOs-2 model therefore follows as $\SI{2.49}{\giga\pascal}$ for the cytoplasm and $\SI{2.24}{\giga\pascal}$ for the nucleus (nucleoplasm and nucleolus combined). Figure \ref{fig:numerical_model_results}(b) shows how the bulk modulus in the range of interest connects to the Poisson ratio for the relatively low static E-modulus of $\SI{1}{\kilo\pascal}$, reported for SaOs-2 cells\cite{holenstein2019relationship}.

The base numerical model of a LM5 cell is then formed by keeping all the material properties of individual cell layers from the calibrated SaOs-2 cell model, but changing the volume of the cell to the volume measured in experiments. In addition, the nucleus-cell volume ratio was changed from $0.434$ to $0.569$, according to previously reported\cite{holenstein2019relationship} values. These changes lead to a decrease in the ACF for the base LM5 cell to $0.031$ whereas the value measured in experiments is $\sim \SI{27}{\percent}$ higher at $0.037$. The decrease in the ACF is due to the increased nucleus-cell volume ratio that decreases the apparent density and the apparent bulk modulus of the cell. In Fig. \ref{fig:numerical_model_results}(c) we explore how a change in the density and bulk modulus across the three layers of a cell through common multipliers could cause the increase in the ACF that would match the experimentally measured ACF of $0.037$ (red line). In Fig. \ref{fig:numerical_model_results}(d), we demonstrate that changing the E-modulus across the cell layers does not directly affect the ACF as long as the bulk modulus and the density are kept at a fixed value.


\section*{Discussion}

Our results reveal that formaldehyde-fixed cells, presumed to have an increased static E-modulus, have an increased ACF compared to untreated cells. Furthermore biological cells treated with the actin depolymerizing drug cytochalasin D (CD), known to reduce the static E-modulus, have a decreased ACF. This simultaneous change of static E-modulus and ACF is particularly interesting when considering that a decrease in the static E-modulus correlates to an increase of the metastatic potential of cancer cells\cite{swaminathan2011mechanical}, especially as the static E-modulus of an object can be linked to its compressibility (eq. \ref{equation_compressibility_K} and \ref{eq:linear_elastic}) and thus to the ACF of the object, as demonstrated by eq. \ref{equation_1d_acoustic_contrast_factor}. This dependence could potentially be used to predict the increased ACF of the static E-modulus increased biological cells and the decreased ACF of static E-modulus decreased biological cells as presented in this study. To test this predictability a hypothesis can be formulated, that the more malignant cells have a decreased ACF and thus interact less with the acoustic field. This hypothesis can be tested with the parental cancer cell line with low metastatic potential SaOs-2 with a static E-modulus of $\SI{0.95}{\kilo\pascal}$ and the high metastatic potential cell line LM5 with a static E-modulus of $\SI{0.8}{\kilo\pascal}$, both static E-moduli were measured with a RT-DC system\cite{holenstein2019relationship}. Contrary to the hypothesis, we found that the difference between the ACFs of the two cell lines with different metastatic potential is only in the range of a few percent, even though the sample size is limited. Nevertheless, this opens up the question if the static E-modulus change is the causation for a change of the ACF or if the change is merely correlated. The broader implication is that forming a hypothesis about dynamic material properties based on known static material properties can lead to imprecise answers and the current literature reflects this uncertainty and caution must be exercised when making these assumptions.  

Cushing \textit{et al.}\cite{cushing2017ultrasound} found that fixing the biological cells with formaldehyde leads to an increase of the compresibility together with a decrease of the density. This leads to an overall decrease of the ACF, which is contrary to the increase of the ACF found in our study. Important to note is that in Cushing \textit{et al.} the measurement of the compressibility was done by speed of sound measurements on a population of biological cells in suspension and not on individual biological cells as was used during this study, which could account for some discrepancies. Furthermore, although the static E-modulus of the cell should have been increased by the fixation, the authors do not try to link the static E-modulus and the dynamically measured compressibility. 

Wang \textit{et al.}\cite{wang2018single} demonstrated another method to measure the density and compressibility of biological cells. Although no statistical analysis was performed, they demonstrated that the compressibility increases with the metastatic potential. The change in the density however did not follow a trend when compared to their metastatic potential. The authors do not specify in which fluid the measurement was performed, therefore it is not possible to precisely calculate the ACF. Assuming PBS as the fluid in which the experiments were conducted, we calculate the ACF of the low and high metastatic cell lines. For these assumed values of the ACF, we could not find a correlation between the metastatic potential and the ACF, in part stemming from the simultaneous change of the compressibility and density. 

\textit{Summa summarum}, there are many gaps in the literature when considering the dynamic properties of biological cells in acoustic fields. Slightly different methods lead to different trends concerning the ACF, and no correlation is attempted to be made between known static material properties and the dynamically measurement data. This is unfortunate as most literature on cell mechanics reports the static E-modulus, which is the standard output for gold standard methods such as the quasi-static AFM.

\begin{table}[h!]
\centering
\footnotesize
\begin{tabular}{l l c c c c c c}
\hline\hline
 & Material of interest & $d [\SI{}{\micro\meter}]$ & $\rho [\SI{}{\gram\per\cubic\centi\meter}]$ & $E [\SI{}{\kilo\pascal}]$ & $G [\SI{}{\kilo\pascal}]$ & $\kappa [\SI{}{\per\tera\pascal}]$ & $\Phi$ / ACF $[1]$ \\\hline
\textbf{a} & Water & - & $0.9966$ & - & - & $444.8$ & - \\ \hline
\textbf{b} & NIH/3T3 & - & $1.079$ & $3-5$ & $1.67$ & $378$ & $0.083$ \\
 & MCF-7 & - & $1.068$ & $0.31-0.6$ & $0.103$ & $422$ & $0.047$ \\
 & HEPG2 & - & $1.087$ & $0.191-0.941$ & $0.0637$ & $428$ & $0.047$ \\
 & HT-29 & - & $1.077$ & $4.09$ & $1.36$ & $404$ & $0.060$ \\
 & MCF-12A & - & $1.068$ & - & - & $377$ & $0.083$ \\
 & RBC & - & $1.099$ & - & - & $331$ & $0.117$ \\
 & Polystyrene & - & $1.050$ & - & - & $216$ & $0.175$ \\ \hline
\textbf{c} & HNC Tu686 & - & $1.025$ & - & - & $405$ & $0.032$ \\
 & HNC 686LN & - & $1.060$ & - & - & $428$ & $0.025$ \\
 & HNC M4e & - & $1.080$ & - & - & $432$ & $0.029$ \\
 & HNC 37B & - & $1.045$ & - & - & $440$ & $0.012$ \\\hline
\textbf{d} & RBC & - & $1.101$ & - & - & $334$ & $0.109$ \\
 & RBCfx & - & $1.091$ & - & - & $356$ & $0.090$ \\
 & WBC & - & $1.054$ & - & - & $393$ & $0.051$ \\
 & WBCfx & - & $1.045$ & - & - & $400$ & $0.042$ \\
 & DU-145 & - & $1.062$ & - & - & $384$ & $0.059$ \\
 & DU-145fx & - & $1.036$ & - & - & $404$ & $0.036$ \\
 & MCF-7 & - & $1.055$ & - & - & $373$ & $0.066$ \\
 & MCF-7fx & - & $1.035$ & - & - & $395$ & $0.043$ \\
 & LU-HNSCC-25 & - & $1.061$ & - & - & $377$ & $0.065$ \\
 & LU-HNSCC-25fx & - & $1.040$ & - & - & $404$ & $0.038$ \\
 & Polystyrene & $5$ & $1.058$ & - & - & $273$ & $0.143$ \\
 & Polystyrene & $7$ & $1.059$ & - & - & $276$ & $0.141$ \\
 & Melamine & $10$ & $1.500$ & - & - & $124$ & $0.363$ \\
 & PMMA & $3$ & $1.184$ & - & - & $173$ & $0.255$ \\\hline
\textbf{e} & MESC2.10 & $13.2$ & - & - & - & - & $0.05$ \\
 & MESC2.10-diff4d & $12.2$ & - & - &- & - & $0.08$ \\\hline
\textbf{f} & Perfluoropentane core & $4.2$ & 1.619 & - & - & $670$ & $(-0.014) - (-0.029)$ \\
& \& cellulose nanofiber shell & & & & & &   \\\hline
\textbf{g} & SaOs-2 & 17.6 & - & 0.95 & - & - & 0.037 \\
 & SaOs-2fx & 15.0& - & - & - & - & 0.050 \\
 & SaOs-2 Cyto. D & 18.9 & - & - & - & - & 0.022 \\
 & LM5 & 15.5 & - & 0.80 & - & - & 0.037 \\
 & Polystyrene & 10.2 & 1.050 & - & - & $250$ & 0.167\\\hline
h & Cytoplasm & $17.6$ & $1.099$ & $0.31$ & - & $402$ & - \\
 & Nucleoplasm & $13.3$ & $0.996$ & $1.78$ & - & $446$ & - \\
 & Nucleolus & $4.3$ & $1.744$ & $5.36$ & - & $446$ & - \\
\hline\hline
\end{tabular}
\caption[ACF of different materials and cell lines, treated and untreated.]{\footnotesize Material properties of eukaryotic cells and some reference inorganic materials. The measurement uncertainty of individual results is omitted for the sake of clarity. Size $d$ generally represents the diameter of a cell, particle, or a part of a cell. $\Phi$ is the mathematical symbol for the acoustic contrast factor (ACF). \\
\textbf{a} Values for water from Karlsen and Bruus\cite{karlsen2015forces} used in numerical models, with the dynamic viscosity of $\SI{0.8538}{\milli\pascal\second}$ and the bulk viscosity of $\SI{2.4}{\milli\pascal\second}$. \\
\textbf{b} NIH/3T3 - fibroblast cell; MCF-7 - breast cancer cell; HEPG2 - liver cancer cell; HT-29 - colon cancer cell; MCF-12A - breast cell; RBC - red blood cell; Hartono \textit{et al.}\cite{hartono2011chip}; the compressibility $\kappa = 1/K$ is measured via acoustic manipulation at $\SI{3.75}{\mega\hertz}$, the other properties are referenced within \cite{hartono2011chip}; density of MCF-12A is assumed. \\
\textbf{c} HNC Tu686, 686LN, M4e, 37B - head and neck cancer cells whereas M4e \& 37B have a higher metastatic potential than Tu686 \& 686LN;  density and compressibility from Wang \textit{et al.}\cite{wang2018single}; acoustic manipulation (data assessed from the graphs) at $\SI{1.91}{\mega\hertz}$. Wang \textit{et al.} do not  report the ACF nor the fluid in which the cells were focused. If the fluid is assumed to be PBS, these are the according ACFs. \\
\textbf{d} `fx' denotes a fixed sample. RBC - red blood cell; WBC - white blood cell; DU-145 - prostate cancer cell; MCF-7 - breast cancer cell; LU-HNSCC-25 - head and neck squamous cancer cell; PMMA - polymethylmethacrylat; Cushing \textit{et al.}\cite{cushing2017ultrasound}; through speed of sound measurements of neutrally buoyant samples at $\SI{3}{\mega\hertz}$; $\Phi$ in PBS ($c=\SI{1508.2}{\meter\per\second}$ and $\rho=\SI{1004.00}{\kilo\gram\per\cubic\meter}$). \\
\textbf{e} MESC2.10 - human embryonic ventral mesencephalic cell; MESC2.10-diff4d - human embryonic ventral mesencephalic cell differentiated in a special medium for 4 days; Augustsson \textit{et al.}\cite{augustsson2010measuring}; the acoustic contrast $\Phi$ is measured via acoustic manipulation at $\SI{1.97}{\mega\hertz}$; we multiplied the average $\Phi$ from \cite{augustsson2010measuring} by a factor of $3$ to match our definition of $\Phi$. \\
\textbf{f} Loskutova \textit{et al.}\cite{loskutova2021measuring} demonstrated a negative ACF, whereas the ACF is dependent on the pressure amplitude. \\
\textbf{g} SaOs-2 \& LM5 - human bone cancer cell line with a lower respectively higher metastatic potential and ACF data generated during this study. Static E-modulus measured using an RT-DC system and taken from Holenstein \textit{et al.}\cite{holenstein2019relationship}. The cell diameter was measured either by a cell counter or by manually determining the size from the videos used to compute the ACF. Cells were focused in the $\SI{742.1}{\kilo\hertz}\pm\SI{2}{\kilo\hertz}$ range. \\
\textbf{h} Data from the calibrated numerical model of the SaOs-2 cell presented in this study.
}
\label{tab:eukaryote_mat_prop}
\end{table}
One major limiting factor to date hindering the direct comparison is that most publications deal with measuring only one material property such as the static E-modulus or the dynamic compressibility. This one parametric view is not sufficient for even the simplest cell mechanics models such as the elasticity theory which requires at least two parameters to describe cell mechanics\cite{nijenhuis2014combining}, as seen in eq. \ref{eq:linear_elastic}. This can then necessitate estimates of further material properties such as the Poisson's ration $\nu$, and whichever ratio is chosen has orders of magnitude impact on the calculated static bulk or E-modulus. This can quickly be seen from the common assumption that cells are incompressible, meaning that $\nu = 0.5$ and would therefore imply an infinitely large bulk modulus as seen from eq. \ref{eq:linear_elastic} and Fig. \ref{fig:numerical_model_results}(\textbf{b}). In order to try to shed some light on the usefulness of static mechanical properties reported for various cell lines, we provide a computational model for a cell in an acoustic field. This model is insofar interesting as it is a multiparametric model where the mechanical properties, such as the E-modulus and the bulk modulus, can be varied independently. The computational model shows on the one hand that the E-modulus can be varied without changing the ACF as seen in Fig. \ref{fig:numerical_model_results}(\textbf{c}). This supports previous findings and the results from our experiments, that a change in E-modulus might be correlated with a change in the ACF but does not necessarily need to be the causation. Therefore the hypothesis that the static E-modulus alone can be used to predict the dynamic material properties must be rejected. To further demonstrate this possible independence, the computational model shows that there is a parameter space where a variation of the apparent bulk modulus and density can result in the same ACF as seen in Fig. \ref{fig:numerical_model_results}\textbf({b}). This finding is further supported by the findings of Wang \textit{et al.} where a change in metastatic potential altered both the density and dynamic compressibility, thus defying a trend between metastatic potential and the ACF. There is another method to measure dynamic properties of a cell (e.g. bulk modulus) - scanning acoustic microscopy\cite{nijenhuis2014combining}, which confirms our predictions of the Poisson's ratio, and confirms that the E-modulus can vary independently of the bulk modulus. This means that most known material properties have an undefined influence on dynamic measurements and prior knowledge cannot be exploited, such as static E-modulus measurements from quasi-static methods such as AFM or RT-DC. This is problematic for applications such as the acoustic focusing and sorting of cells in liquid biopsies, potential cancer treatment methods such as oncotripsy, ultrasound neuromodulation\cite{rabut2020ultrasound, tufail2010transcranial} and sonogenetics\cite{ibsen2015sonogenetics}. This indicates an unmet need for dynamic material properties which in parts could be potentially alleviated by AF.


\section*{Methods}\label{sec:methods}

\subsection*{Theoretical Background: Acoustic scattering and streaming}

The presented work deals with objects in fluids. In order to predict the motion of the objects, the fundamental equations of the motion of the fluids need to be introduced. The motion of a viscous fluid is governed by the compressible Navier-Stokes equations
\begin{equation}
    \rho \left[ \frac{\partial \underline{v}}{\partial t} + ( \underline{v} \cdot \vect{\nabla} ) \underline{v} \right] = - \grad p + \eta \lapl \underline{v} + \para{ \bvis + \frac{\eta}{3}} \grad \para{\divg \underline{v}} \label{al:NS} ,
\end{equation}
and the continuity equation
\begin{equation}
    \frac{\partial \rho}{\partial t} = -\rho \divg \underline{v} \label{al:cont} ,
\end{equation}
with the velocity $\underline{v}$, pressure $p$, the dynamic viscosity $\eta$ and the bulk viscosity $\eta_{\mathrm{B}}$. As the fluid is assumed to be barotropic, the density $\rho$ is assumed to be a function of pressure $p$ only $\rho= \rho(p)$.

The equations are linearized using the perturbation approach \cite{bruus2012perturbation}. Accordingly, the physical fields are expanded in a series, $\square = \square_0 + \square_1 + \square_2 + \dots$, where $\square$ represents the field, while the subscript denotes the respective order.

\subsubsection*{First-order (acoustic) problem}
For a quiescent fluid at the zeroth order ($\underline{v}_0 = \underline{0}$), the substitution of the perturbed fields into the governing equations yields the following set of linear first-order equations,
\begin{align}
\rho_0 \frac{\partial \underline{v}_1}{\partial t} &= - \grad p_1 + \eta \lapl \underline{v}_1 + \para{ \bvis + \frac{\eta}{3}} \grad \para{\divg \underline{v}_1} \label{al:eq1} ,\\
\frac{\partial \rho_1}{\partial t} &= -\rho_0 \divg \underline{v}_1 \label{al:eq2} ,
\end{align}
with the equilibrium density $\rho_0$. The equation of state,
\begin{equation}
\rho_1 = \frac{1}{c_0^2} p_1,
\label{eq:state}
\end{equation}
is connecting the first-order density with the first-order pressure through the speed of sound in the fluid $c_0$. The first-order fields are assumed to have a harmonic time-dependency with the factor $\euler^{\imag \omega t}$, with the angular frequency $\omega = 2 \pi f$.

The acoustic fields, comprised of the velocity $\underline{v}_1$ and pressure $p_1$, are assumed to be the sums of background fields (bg) and scattered fields (sc), namely $(\,)_1=(\,)_{1\mathrm{bg}}+(\,)_{1\mathrm{sc}}$. We assume a one-dimensional plane standing wave along the $z$-direction of the cylindrical coordinate system. The background velocity field is set to

\begin{equation}
\underline{v}_{1\mathrm{bg}} = \mathrm{Re} \bkt{\frac{\velpot_{\na}}{2}\imag k \para{\euler^{\imag k z }-\euler^{-\imag k z }}\euler^{\imag \omega t }} \underline{e}_{z},
\label{eq:bckg_velocity}
\end{equation}
with the corresponding velocity potential amplitude
\begin{equation}
\velpot_{\na}= -\frac{p_{\na}}{\imag \omega \rho_0 + \para{\bvis+\frac{4}{3} \eta}k^2},
\label{eq:vel_potential}
\end{equation}
with pressure amplitude $p_{\na}$, and the wavenumber
\begin{equation}
    k = \frac{\omega}{\cfl} - \alpha \imag,
    \label{eq:wavenumber}
\end{equation}
with the attenuation coefficient for viscous fluids \cite{blackstock2001fundamentals}
\begin{equation}
    \alpha = \frac{\omega^2}{2 \cfl^3 \rho_0} \left( \eta_{\mathrm{B}} + \frac{4}{3} \eta \right).
\end{equation}

\subsubsection*{Second-order (streaming) problem}
Applying the perturbation theory up to second order to the governing equations, together with taking the time average $\langle \square \rangle = \frac{1}{T}\int_T \square \mathrm{d}t $ over an oscillation period $T=1/f$, results in the equations of acoustic streaming \cite{doinikov1994acoustic},
\begin{align}
    \grad \avg{p_2}- \eta \lapl \avg{\underline{v}_2} -\para{\bvis+\frac{\eta}{3}} \grad \para{\divg \avg{\underline{v}_2} } &= - \rho_0 \divg \avg{\underline{v}_1 \underline{v}_1}, \label{al:PartStramIII} \\
    \rho_0 \divg\avg{\underline{v}_2} &= -\divg \avg{\rho_1 \underline{v}_1}. \label{al:PartStramIV}
\end{align}

\subsection*{Theoretical Background: Acoustic contrast factor}
A method to measure the acoustic contrast factor of a biological cell is to subject the biological cell to a standing pressure wave within a fluid cavity. The standing pressure wave is generated by exciting a piezoelectric transducer (PT) glued onto an acoustofluidic device, which transmits the vibration of the PT into the fluid cavity where resonance is established. These vibrations, when assuming a spherical particle with a radius much smaller than the acoustic wavelength in an inviscid fluid, give rise to a potential commonly defined as the Gor'kov potential $U$\cite{gor1962forces}

\begin{equation}
    U = \frac{4 \pi}{3} r^3 \biggl[ f_{1}(\tilde{\kappa}) \frac{1}{2 \rho_{0} c_{0}^2} \langle p_{1 \mathrm{bg}}^2 \rangle - f_{2}(\tilde{\rho}) \frac{3}{4} \rho_{0} \langle \underline{v}_{1 \mathrm{bg}}^2 \rangle \biggr],
    \label{equation_gorkov_potential}
\end{equation}
where $r$ is the radius of the particle, $\langle p_{1 \mathrm{bg}}^2 \rangle$ the first order time averaged square of the incident acoustic pressure and $\langle \underline{v}_{1 \mathrm{bg}}^2 \rangle$ the first order time averaged square of the incident acoustic velocity. Furthermore, the monopole scattering coefficient $f_{1}(\tilde{\kappa})$ which is related to the relative compressibility between the particle and the medium and the dipole scattering coefficient $f_{2}(\tilde{\rho})$ which is related to the relative density between the particle and the medium
\begin{equation}
    f_{1}(\tilde{\kappa}) = 1 - \tilde{\kappa} ; \\ \tilde{\kappa} = \frac{\kappa_{p}}{\kappa_{0}},
    \label{equation_f1}
\end{equation}

\begin{equation}
    f_{2}(\tilde{\rho}) = 2 \frac{\tilde{\rho} - 1}{2 \tilde{\rho} + 1}; \\ \tilde{\rho} = \frac{\rho_{p}}{\rho_{0}},
    \label{equation_f2}
\end{equation}
where $\kappa_p$ is the compressibility of a particle, $\kappa_0$ is the compressibility of the fluid and $\rho_p$ is the density of the particle.  The compressibility is in this case defined as the the inverse of the bulk modulus $K$

\begin{equation}
    \kappa = \frac{1}{K}.
    \label{equation_compressibility_K}
\end{equation}
The bulk modulus is one parameter needed for the linear elastic description of cell mechanics and is given by 

\begin{equation}
    K = \frac{E}{3(1-2\nu)}
    \label{eq:linear_elastic}
\end{equation}
for a given static E-modulus and Poisson's ratio $\nu$. For fluids, bulk modulus follows as $\rho_0 c_0^2$, and the compressibility as $\kappa_0 = 1/ ( \rho_0 c_0^2 )$. The acoustic radiation force $\underline{F}_{ARF}$ is commonly approximated as the negative gradient of the Gor'kov potential $U$

\begin{equation}
  \underline{F}_\mathrm{ARF} = -\boldsymbol{\nabla} U.
  \label{equation_f_rad}
\end{equation}
Given an inviscid standing pressure wave defined through eq. \ref{eq:bckg_velocity} and assuming $\eta= \eta_{\mathrm{B}} = 0$, $\underline{F}_{ARF}$ can be simplified to a one-dimensional ARF in the $z$ direction

\begin{equation}
    F^{1D}_\mathrm{rad}  = 4 \pi r^3 E_{ac} k_z sin\left(2k_z \left(z+ \frac{w}{2}\right)\right) \Phi(\tilde{\kappa},\tilde{\rho}).
    \label{equation_1d_gorkov_potential}
\end{equation}
$k_z = \frac{\omega}{c_0}$ is the wavenumber in an inviscid fluid, $w$ is the width of the channel, $\sin(2k_z (z + w/2))=1$ for the maximal force, $E_{ac}$ is the acoustic energy density, the direction of $z$ is shown in Fig. \ref{fig:numerical_model} and $\Phi$ is the acoustic contrast factor 

\begin{equation}
    \Phi(\tilde{\kappa},\tilde{\rho})=\frac{1}{3}f_1(\tilde{\kappa})+\frac{1}{2}f_2(\tilde{\rho}).
    \label{equation_1d_acoustic_contrast_factor}
\end{equation}
which is a dynamic quantity related to the scattering of pressure waves on objects and is related to the relative density and compressibility between a fluid and an object in the fluid. A table of relevant $\Phi$ values is given in table \ref{tab:eukaryote_mat_prop}, whereas objects with a positive ACF migrate to pressure nodes as seen in Fig. \ref{fig:setup} and objects with a negative ACF migrate to pressure anti-nodes. The drag force acting on a particle moving through a fluid is given by the Stokes' drag\cite{nyborg1965acoustic, lighthill1978acoustic}  

\begin{equation}
    \underline{F}_{str}= - 6\pi \eta r  \underline{v}_{p},
    \label{equation_stokes_drag}
\end{equation}
where $\underline{v}_{p}$ is the velocity of the particle. Neglecting the particle inertia, the particle velocity $v_z$ in the $z$ direction can be calculated by balancing the ARF with the Stokes' drag\cite{bruus2012arf}

\begin{equation}
    v_{z}=\frac{2\Phi}{3\eta}r^2 k_z E_{ac} \sin(2k_z (z+\frac{w}{2})).
\label{equation_balance}
\end{equation}
With $v_z=\frac{\partial{z_p}}{\partial{t}}$ and given a particle at starting position $z_0$ at time $t=0$, the transverse path $z_p(z_0,t)$ can be calculated\cite{bruus2012arf}

\begin{equation}
z_p(z_0,t)=\frac{1}{k_z}\arctan\left\{\tan\left[k_z (z_0+\frac{w}{2})\right]\exp(\frac{4\Phi}{3\eta}(k_z r)^2 E_{ac} t)\right\}-\frac{w}{2}.
\label{equation_trajectory}
\end{equation}
The resonance frequency for the $n$-th ultrasonic resonance mode of a one-dimensional standing wave with hard wall boundary conditions is\cite{bruus2012acoustofluidics}

\begin{equation}
    f_{res}^{1D}= \frac{c_{0} n}{2w} = \frac{c_{0}}{\lambda_n},
    \label{equation_nth_mode}
\end{equation}
where $w$ is the width of the channel and $\lambda_n$ is the wavelength. For $n = 1$ or $n=2$, the wavelength is equal to $2w$ or $w$ corresponding to the $\lambda/2$ mode or $\lambda$ mode, respectively. 
 

\subsection*{Setup}

The setup is analogous to the one used in Harshbarger \textit{et al}.\cite{harshbarger2022optical} and is comprised of a function generator (AFG-2225, GW Instek) which is connected to a computer running a code which allowed for the one click change of the frequency. The signal from the function generator is amplified (325LA Linear Power Amplifier, Electronics \& Innovation) and monitored using an oscilloscope (UTD2025CL, UNI-T). The amplified signal is passed on to the piezoelectric transducer (PT). The vertical setup seen in Fig. \ref{fig:setup}(\textbf{a}) is built from parts from the THORLABS Cerna® series. The video feed was generated with a 10x objective (M Plan Apo 10x / 0.28, Mitutoyu) and a uEye camera (UI-3160CP Rev. 2.1, iDS, 1920 x 1200 pixels, 60 fps). To positively identify the fluorescent PS in the solution, a blue LED was used. In addition, there was a backlight, shining through the backside of the device, which resulted in brightfield image, but also an increased fluorescent signal of the PS particles. The sample flow was controlled manually whereas the $\SI{1}{ml}$ syringe was placed vertically in a holder located above the device in order to avoid sedimentation of the PS particles and the cells in the tubing and the device. This vertical orientation is unlike the horizontal orientation presented in\cite{hartono2011chip, wang2018single, cushing2017ultrasound, augustsson2010measuring}. This vertical orientation is only advisable if the density of the objects in the fluid are in a similar range as the density of the fluid, as is the case for biological cells and PS particles in PBS. 

\subsection*{Glass-silicon-glass device}

The glass-silicon-glass device, seen in Fig. \ref{fig:setup}(\textbf{b}), is produced by bonding a $\SI{500}{\mu\meter}$ thick glass wafer to a $\SI{200}{\mu\meter}$ thick silicon wafer. The fluidic channel is patterned onto the exposed silicon wafer using photolithography (resist:  S1828, Shipley, 4'000 $\mathrm{rpm}$; developer: AZ351B, Microchemicals). The full thickness of the silicon wafer is etched away with an inductively coupled plasma (ICP) deep reactive ion etching (DRIE) machine (Estrellas, Oxford instruments). Following the etching, a $\SI{700}{\micro\meter}$ thick glass wafer is anodically bonded onto the exposed silicon wafer. The wafer is diced into individual devices with a wafer saw (DAD3221, Disco corporation). Fused silica capillaries ($164\pm\SI{6}{\micro\meter}$ outer diameter, $100\pm\SI{6}{\micro\meter}$ inner diameter, Molex) are inserted into the inlets and outlets of the device to create a fluidic connection. The capillaries are fixed with a two-component glue (5 Minute Epoxy, Devcon). Piezoelectric transducers ($\SI{10}{\milli\meter}$ length, $\SI{2}{\milli\meter}$ width, $\SI{1}{\milli\meter}$ thickness, Pz26, Meggitt Ferroperm) are glued on using an electrically conductive Epoxy glue (H20E, Epoxy Technology). Copper cables ($\SI{0.15}{\milli\meter}$ diameter) are attached to the piezoelectric transducers and the electrical connection was established with an electrically conductive silver paste. The final dimensions of the devices is given by: device: 50 mm length, 12 mm width, 1.4 mm thickness; focusing channel: 15 mm length, 1 mm width, 0.2 mm height as seen in Fig. \ref{fig:setup}c. The 3 inlets results in the cells already being slightly prefocused close to the walls before the acoustics is turned on. This makes the prefocusing process in a $\lambda$ mode efficient, if even needed.

\subsection*{Polystyrene particles}
Green fluorescent polystyrene (PS) particles (microParticles GmbH, Germany) with diameters of $10.23\pm \SI{0.13}{\micro\meter}$ and 2.5 w/v\% were used for all experiments. 

\subsection*{Cell culture}
Two cell lines were used, the low metastatic potential parental cell line SaOs-2\cite{fogh1975new} and the highly metastatic cell line LM5. The creation of the LM5 cell line is detailed in\cite{jia1999nude}. In brief, Sarcoma Osteogenic (SaOs)-2 cells were isolated from the primary osteosarcoma of an 11 year old Caucasian girl and were subsequently injected into the tail vein of a nude mouse. After 6 months the mouse was sacrificed and the metastatic lesion was removed from the lung of the mouse. The cells isolated from the lung lesion were again injected into the tail of a new nude mouse. This process was repeated 4 times resulting in the lung metastasis 5 (LM5) cell line. The anatomical origin of the primary tumor from which the SaOs-2 cells were extracted is not known. All cell lines are kept at the standard 37\textdegree C and 5\% $CO_2$ and 95\% air. The cell media used is DMEM - F12 Ham (D8437, Sigma) supplemented with 10\% fetal bovine serum (10270106, Thermo) and 1\% P/S. The cells were passaged when 60\% confluency was reached. The diameter of the cells was measured by the CellDrop BF, DeNovix after removal from the cell culture flask or directly in the video. The results section always indicated how the cell size was measured. 

\subsection*{Trajectory measurement protocol}
The cells were removed from a T75 culture flask using Trypsin and centrifuged for $\SI{5}{min}$ at $\SI{300}{g}$. After spinning down the cells, the excess culture medium was removed. If needed, the cells were treated by either:

\begin{itemize}
    \item fixing the biological cells with $\SI{1}{\milli\liter}$ of a 4\% formaldehyde solution (ROTI Histofix, CARL ROTH) for 30 minutes at room temperature, whereas formaldehyde functions as a protein crosslinker. Afterwards, the cells were washed twice with PBS to remove the fixative. Or
    \item with $\SI{300}{\micro\liter}$ cytochalasin D, which is an inhibitor of actin filament polymerization, whereas the actin filaments play a fundamental role in controlling cell shape and mechanical properties\cite{wakatsuki2001effects}. A cytochalasin D (Sigma-Aldrich) stock solution of $\SI{5}{\milli\gram\per\milli\liter}$ was diluted down to $\SI{333}{\nano\mole\per\milli\liter}$. The cells were incubated for 30 minutes at room temperature
\end{itemize}
or were left untreated. $\SI{25}{\micro\liter}$ of the $10.23\pm \SI{0.13}{\micro\meter}$ PS particle solution were added to the non-treated or treated cells. PBS was added to bring the final volume to $\SI{1}{\milli\liter}$. The PS particles and biological solution was loaded into a $\SI{1}{\milli\liter}$ syringe, and sequentially flown into a pre-characterised acoustofluidic device. The flow was turned off during the recording of the video sequence. Once a video sequence was recorded, a fresh suspension of PS particles and biological cells were flown into the device. The default frequency continuously applied was $f=\SI{1.42}{MHz}\pm\SI{2}{\kilo\hertz}$ as this corresponds to a $\lambda$ mode, i.e. two pressure nodes in the device. This allows for a prealignment of the PS particles and biological cells, and reduces the measurement error due to a minimized variability of the starting position. The video recording was started, and the frequency was changed from a $\lambda$ to a $\lambda/2$ mode found at $f = \SI{742.1}{\kilo\hertz}\pm\SI{2}{\kilo\hertz}$. 

\subsection*{Image analysis}
The generated videos were analysed using Fiji\cite{schindelin2012fiji}. The videos were converted to an Image Sequence. This allows for the selection of the Image Sequence starting from the first movement of the PS particles and cells in solution until the last frame where movement could be seen. The particles are identified and separated into either PS or cells by using one of two methods: The machine learning software ilastik\cite{berg2019} can be used. This allows for the automatic detection and classification of particles of interest, but requires a training set. For videos with low particle numbers, it is faster to track the particles by hand by going through the Image Sequence frame by frame and marking the particle of interest. The particle tracks are calculated using the Fiji Plugin TrackMate. 

\subsection*{Data analysis}
The biological cell and PS particle trajectories from TrackMate are imported into a custom MATLAB script. Importing the known material parameters, seen in table \ref{tab:eukaryote_mat_prop}, of the PS particles and PBS and using the trajectory eq. \ref{equation_trajectory}, the acoustic energy density $E_{ac}$ within the system can be calculated. Using the extracted $E_{ac}$, the ACF of the biological cells in the fluid suspension can be calculated. 

\subsection*{Numerical model}
In the scope of our study, we developed a finite element method (FEM) model of an eukaryotic mammalian cell whereas the structure is based on the model of Heyden and Ortiz\cite{heyden2016oncotripsy}, consisting of nucleolus, nucleoplasm, and cytoplasm, which are modeled as elastic solids. A graphical summary of the simplified model of the cell is given in Figure \ref{fig:numerical_model}.

The FEM model is built in COMSOL Multiphysics\textregistered~ v. 5.6 \cite{comsol}, following and building upon the general approach of Baasch \textit{et al.}\cite{baasch2019acoustic}. The perturbed equations \ref{al:eq1}, \ref{al:eq2} and \ref{eq:state} of acoustic scattering and equations \ref{al:PartStramIII} and \ref{al:PartStramIV} of acoustic streaming are solved consecutively. The scattering is solved in a frequency domain study, while the streaming is computed in a stationary study, using the results of the scattering study as source terms in equations \ref{al:PartStramIII} and \ref{al:PartStramIV}. The solid domains are modeled via the Solid Mechanics interface, where the appropriate material models are defined. The fluids in the frequency domain study are modeled with a Thermoviscous Acoustics interface with the adiabatic formulation, and afterwards, in the stationary study, with a Creeping Flow interface.

At the first order, we impose the continuity of velocity and stress at the fluid-solid interface. The fluid is modelled as unbounded, and far away from the particle the first-order fields converge to the background fields, defined by eq. \ref{eq:bckg_velocity}.
At the second order, the no-slip boundary condition is imposed on the Lagrangian velocity of a fluid at the fluid-solid interface, in order to compensate for the oscillations of the interface at the first order. The Lagrangian velocity is defined as the summation of the Eulerian streaming velocity $\left< \underline{v}_2 \right>$ and the Stokes drift \cite{andrews1978exact, buhler2014waves}
\begin{equation}
    \underline{v}_{\mathrm{SD}} = \avg{ \left( \int{\underline{v}_1}\mathrm{d}t \bcdot \grad \right) \underline{v}_1},
\end{equation}
which consequently translates into the boundary condition
\begin{equation}
\avg{\underline{v}_2} = - \underline{v}_{\mathrm{SD}} \quad \text{at the interface}.
\end{equation}
The streaming due to the attenuation of the background standing wave in the absence of the particle is negligible \cite{doinikov1994acoustic,baasch2019acoustic} and was neglected in the present study.

The FEM model is axisymmetric to limit the computational effort. In the frequency domain study, the fluid domain surrounding the cell is surrounded by a perfectly-matched layer (PML) that absorbs any incoming waves, in order to avoid any influence of the outer wall on the cell. For the steady study, where the acoustic microstreaming is computed, the PML is replaced by a no-slip boundary condition, and the wall effects are avoided by ensuring that the fluid domain is large enough.


\bibliography{bibliography}


\section*{Acknowledgements}

We would like to thank the Balgrist Foundation and ETH Zurich for their most generous financial support. 

\section*{Author contributions statement}

C.H. conceived the experiment, designed the setup and produced the microfluidic devices; C.H., D.B. and A.V. conducted the experiments; A.P. created the numerical model; C.H. and A.P. analyzed the results and wrote the draft of the manuscript; J.S., J.D. and U.S. reviewed the manuscript. 

\section*{Competing interests}

The authors declare no competing interests.

\end{document}